# ON THE CORROSION RESISTANCE OF POROUS ELECTROPLATED ZINC COATINGS IN DIFFERENT CORROSIVE MEDIA

Y. Hamlaoui[1], F. Pedraza[2*], L. Tifouti[3]

1. Institut des Sciences et Sciences de l'Ingénieur. Centre Universitaire de Souk-Ahras. BP 1553. 41000 Souk-Ahras. Algeria.

2. Laboratoire d'Etudes des Matériaux en Milieux Agressifs (LEMMA). Pôle Sciences et Technologie. Université de La Rochelle. Avenue Michel Crépeau. 17042 La Rochelle Cedex 1. France.

3. Laboratoire de Génie de l'Environnement. Université Badji Mokhtar. BP 1223. 23020. El Hadjar-Annaba. Algeria.

* Corresponding author : Tel. : +33(0)546458297. Fax : +33(0)546457272.

e-mail : fpedraza@univ-lr.fr

**Abstract**

The corrosion resistance of an electroplated (EP) Zn coating whose surface was chemically etched to produce surface defects (pores) is investigated in this work. Impedance and DC polarisation measururements were employed to study the behaviour of such coating in various corrosive media (NaCl, NaOH and rain water). Four different faradaic relaxation processes were clearly revealed in different NaCl concentrations (from 0.1M to 1M). In the most concentrated solutions at least three relaxation processes at low frequencies (LF) appeared and were related to zinc deposition and dissolution. At lower concentrations and depending on the pH, only one process was observed. The charge transfer resistance ($R_{ct}$) and the corrosion current ($I_{corr}$) were practically stable in the pH range 5 to 10. In deaerated NaCl 0.1M, the EIS diagrams showed two time-constants at very close frequencies. From the EIS diagrams the porous nature of the coating was highlighted and showed that the dissolution mechanisms occurred at the base of the pores.

## 1. INTRODUCTION

Zinc is widely employed as a sacrificial protective coating to prevent corrosion of steel. Such coatings can be deposited by different ways like electroplating, hot dip galvanising or spray. When applied by electroplating, the coating is pure zinc, which is oxidized into a zinc oxide layer to a depth of more than 100 Å upon exposure to air [1] thereby reducing the corrosion rate. However, the corrosion performance of the zinc-based coatings is mostly evaluated without considering the effect of surface defects [e.g. 2-6]. In contrast, Deslouis et al. [7] showed that corrosion of zinc in aerated $Na_2SO_4$ solutions occurred essentially at the base of the pores of the coating. Progress on the anodic dissolution gave rise to four loops with decreasing frequency [8]. Based on this, further insight on Zn dissolution in sulphate solutions was again provided by Cachet et al. [9], who described a reaction model in which three paths associated with three adsorbed intermediates ($Zn_{ad}^+$, $Zn_{ad}^{2+}$ and $ZnOH_{ad}$) were identified depending on the initial surface condition. To identify these steps, most of the works required either the use of low salt concentrations (< 0.1M) to reveal the slowest reactions using electrochemical impedance spectroscopy (EIS) or the use of high salt concentrations (1M) to quantify the corrosion process by D.C. polarization measurements.

In contrast to the above mentioned studies, the purpose of this paper is to report on the results of impedance spectroscopy in combination with D.C. polarisation measurements at various

NaCl concentrations, NaOH and rain water to study the corrosion behaviour of an electroplated zinc coating on steel to which some porosity has been induced by etching the surface with weak HCl solutions

## 2. EXPERIMENTAL PROCEDURE

### 2.1. Sample preparation and coatings

The substrate was a non alloyed steel S235JR (nominal composition according to EN10025-2:2004: Fe- 0.17C - 0.641Mn –0.040S –0.012N –0.55Cu -0.14Si and (<0.035) P, wt%). The EP coatings were obtained from an aerated cyanide-free bath [17-21 g.L$^{-1}$ ZnSO$_4$, 125-135 g.L$^{-1}$ NaOH] at 23 ºC by imposing 2 A.dm$^{-2}$ for 30 min. Thereafter, the coatings were thoroughly rinsed in ethanol in an ultrasonic bath. In order to create the defects (pores) at the surface of the coatings, the samples were immersed for 15 seconds in 10 vol% HCl solutions at room temperature. then thoroughly rinsed in deionised water, ethanol and finally dryed with hot air.

### 2. 2. Corrosion tests and experimental set up

Samples of 4x4x0.05 cm were cut from the electroplated sheets and most of the surface was protected with an adhesive film to leave a 1 cm$^2$ surface in contact with the corrosive medium. All the corrosion tests were normally repeated two or three times, checking that they presented reasonable reproducibility. The corrosion tests were carried out at room temperature by magnetic stirring the solutions to obtain a slight vortex of the electrolyte. The corrosive media consisted of 0.1, 0.5 and 1 M NaCl, of 1M NaOH and of natural rain water. This latter was collected in the trial site of DRA Annaba (Algeria). The chemical composition is given in Table 1.

The electrochemical experimental set-up was composed of a classic three electrode cell using a platinum grid as counter electrode and a saturated calomel electrode (SCE) as the reference one, the coated samples being connected to the working electrode. The measurements were carried out using a potentiostat/galvanostat EGG 273A coupled to a frequency response analyser (FRA) EGG 1025. The impedance data were obtained at the corrosion potential ($E_{corr}$) between 100 kHz and 100 mHz at 10 mV as the applied sinusoidal perturbation. The Tafel polarisation curves were obtained at a scanning rate of 60mV/ min between ±250mV compared to the corrosion potential ($E_{corr}$). The experiments were monitored using the software EGG M352 and Powersine and the EIS results were curve-fitted using software EQUIVCRT designed by Holland Researcher B. Boukamp [10].

The characterisation of the coatings and corrosion products was accomplished by X-ray diffraction (XRD) in a Bruker AXS D8-Advanced diffractometer using Cu-$K_\alpha$ radiation ($\lambda$ = 1.5406 nm) at a scan rate of 0.04°.sec$^{-1}$ in the Bragg-Brentano configuration. In addition, imaging and elemental analyses were performed in FEI Quanta 200 Environmental Scanning Electron Microscope (SEM) coupled to Energy Dispersive Analysis (EDS).

## 3. RESULTS AND DISCUSSION

Figure 1 gathers different SEM images that show that the Zn coatings covered homogeneously the surface of the steel even after etching in 10 vol%HCl (Fig. 1a). The average diameter of the pores was of about 10 μm. When larger, the underlying steel substrate was readily observed as the thickness of the Zn coating was of 10 μm. The EDS microanalyses hence revealed variable amounts of Fe when spotting in the pores.

### 3. 1. Coating behaviour in NaCl solution

### 3.1.1. Effects of concentration and of pH

At least three polarisation tests were carried out for each concentration and pH (Figure 2a). The average values are given in Table 2. As indicated by the Tafel slopes, the corrosion rate in the aerated NaCl medium was found to be under cathodic control (oxygen diffusion). From the shape of the polarization curves obtained on the untreated and treated samples, it can be readily observed that the corrosion rate in the aerated NaCl medium is under cathodic control (oxygen diffusion). Therefore, the corrosion current $i_{corr}$ was calculated using the Stern and Geary equation "$i_{corr} = \beta_a / 2.3R_p$" [11]. In addition, for the 0.1 and 0.5M curves, the corrosion potential was close to the mixed potential of Fe/Zn (-990 mV/SCE) [4]. However, in the 1M curves, the corrosion potential shifted towards more cathodic values suggesting further degradation of the coating than in 0.1 and 0.5M solutions In addition, the corrosion current ($j_{corr}$) also increased with increasing the NaCl concentration, hence the corrosion rate.

Figure 2(b) shows the cyclic polarisation curves obtained in 0.1 M and 1M NaCl. The cyclic hysteresis during the reverse scan was more pronounced in the 1M solutions than in the 0.1M ones, as a result of the modification of the surface of the electrode during the anodic polarisation. Moreover, the evolution of the specific surface indicated an important degradation of the coating. After the polarisation tests in 1M, the XRD patterns (Figure 3) revealed $Zn_6(OH)_8Cl_2.H_2O$ (simonkolleite) and zinc hydroxides as the major corrosive species. However, no iron derivatives were detected suggesting that the either the corrosion products did not reach the steel substrate, as reported by Yadav et al. [6]; or that they were in very low amount to be detected by XRD (see Fig. 1).

The D.C. methods developed for detecting porosity based on the anodic current (or potential) measurements [13-15], or on the mixed potential theory [16] are not applicable to electroplated zinc coatings on steel. Indeed, the former method requires the substrate being electrochemically active with the coating remaining inactive; while the latter needs a fully anodic and a fully cathodic behaviour of the less and f the more noble metals of the couple, respectively. Hence, A.C. impedance methods were proposed for the determination of porosity by examination of the capacity loops in the Nyquist diagrams [17]. With this in mind, EIS measurements were then carried out in different NaCl concentrations. First, the EP zinc coatings were immersed for several days in 0.1M NaCl solutions to ensure that no pores developed upon immersion or polarisation (Figure 4.) The Nyquist diagrams indicated that corrosion of untreated zinc was controlled by a charge transfer reaction, with an increase of the polarisation resistance during immersion. This could be associated with the build-up of zinc corrosion products, which gave rise to an apparent passivation effect because of the accumulation of corrosion products. Although the second one not well defined, two capacitive loops were observed at the beginning of immersion. They tended to disappear with time to result in a straight line at low frequencies due to the development of a Warburg resistance. This was indicative of a significant dissolution of the as-electroplated zinc coating.

The impedance diagram of the etched coatings immersed in a 0.1M NaCl solution showed two relaxation times [Figure 5, Table 3]. The first one corresponded to a high frequency (HF) capacitive loop flattened and deformed on its left side, which is typical of a charge transfer whereas the second one could be ascribed to an inductive loop in the capacitive plane. When the concentration was increased to 0.5M, a new loop weakly initiated and became well defined at 1M NaCl. Therefore, the diagram at LF is composed of a capacitive loop between two inductive loops. In the literature, only one model was proposed to describe the process of dissolution of pure zinc in $Na_2SO_4$ [7] or in NaCl [8,9,18]. The model highlighted two processes in parallel with the presence of two adsorbed species (see reaction pathway on Figure 5). Thus, the EIS diagrams would comprise three low frequency (LF) relaxation processes (θ). In the present study three relaxation processes are revealed in the most concentrated solution (1M). According to Cachet et al. [8] the deformation of the left part of the capacitive loop HF is related to the existence of pores on the surface of the electrode with the substrate surface being inactive. Therefore the four relaxation processes observed in the EIS diagrams can be allotted to: (i) and (ii) the capacitive and inductive loops at HF related to the charge transfer and to the presence of $Zn^I$ (intermediate species), respectively; and (iii) and (iv) the capacitive and inductive loops at LF related to the precipitation and $Zn^{II}$ migration by diffusion and to the dissolution of the oxide layer with time, respectively.

However, one has to consider that the increase of NaCl concentration from 0.1 to 1M brought about a shift of pH from 6 to 7. Therefore, the influence of pH of the solution on the number of the relaxation processes observed on the EIS diagrams was also studied at 0.1M (Figure 7 and Tables 2 and 3. When passing from acidic to alkaline pH no striking difference could be observed. Indeed, the disappearance of the two relaxation processes could arise the precipitation and $Zn^{II}$ migration by diffusion and to the dissolution of the oxide layer with time. One of the phenomena that could explain these features is related to the formation and transformation of the $Zn^{II}$ species occurring rapidly. Therefore the surface is weakly covered - ($\theta_1$) is low- and require frequency sweeps lower than 0.1 Hz to reveal its relaxation process. The other likely phenomena could be associated with a temporary passivation (chemical stability) of the oxide /hydroxide layer formed on Zn, which could slow down the fourth relaxation process. The weak appearance of an inductive loop LF in the EIS diagrams obtained in the 0.5M concentration confirmed these assumptions. Moreover, the low values of $C_{dl}$ and the stability of the $R_{ct}$ values obtained at pH 5 and 10 indicated the establishment of a chemical quasi-equilibrium of the corrosion products (temporary passivation) in agreement with the Pourbaix diagrams [19]. However, this layer started to lose its stability at pH=11. By increasing the NaCl concentration from 0.1 to 0.5 and from 0.1 to 1 M, the charge transfer resistance ($R_{ct}$) lost 21 and 44% of its value, respectively. However, the ($R_{ct} \times I_{corr}$) product remained practically constant (20 mV), thus confirming that the corrosion rate increased with the increase of the NaCl concentration.

### 3.1.2. Influence of oxygen (deaerated and aerated NaCl)

With the aim of studying the influence of oxygen, a series of tests was carried out in aerated and deaerated 0.1M NaCl solutions (Figure 8, Table 4). The oxygen concentration was measured by a standard oxymeter Z621 Consort apparatus and showed between 8 to 10 ppm and 1 to 2 ppm in the aerated and deaerated solutions, respectively. In deaerated solutions, the Nyquist diagrams showed two overlapped capacitive loops. Similarly, the Bode diagram exhibited two very close time-constants. This behaviour could be explained by a fast filling of the pores with corrosion products mainly composed of $Zn(OH)_2$, which could be later converted into the oxide according to equation <1> [20]:

$$Zn^{2+} + 2\ OH^- \rightarrow Zn(OH)_2 \rightarrow ZnO + H_2O \qquad <1>$$

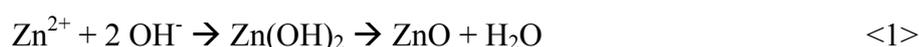

ZnO is considered a semiconductor, hence it may decrease the corrosion protection offered by corrosion products [21]. However, the presence of this species cannot alter the number of time constants because it only appears at the end of the polarisation process (dehydration of the hydroxide).

In addition, the decrease of the oxygen content in the solution slowed down the cathodic reaction (oxygen reduction) and slightly decreased the pH on the surface of the electrode. The variation of the pH near the electrode in the aerated and deaerated media was calculated according to equation <2> and assuming the oxygen reduction as the only cathodic reaction:

In equilibrium: $i_c = \dfrac{[O_2]_b n D_{O_2} F}{\delta} = \dfrac{([OH^-]_s - [OH^-]_b) D_{OH^-} F}{\delta}$ <2>

In this expression $D_{O_2}$ and $D_{OH^-}$ are the diffusion coefficients of oxygen and hydroxide, respectively; $\delta$ is the thickness of the Nernst layer, $[O_2]_b$ and $[OH^-]_b$ are the concentrations of oxygen and hydroxide in the bulk solution, respectively, $[OH^-]_s$ is the hydroxide concentration at the surface of the electrode, F is the Faraday number and n is the number of exchanged electrons (it can be 2 or 4). As the hydroxide concentration in the bulk solution was very low, it was considered that $[OH^-]_b = 0$. By replacing this in equation <2> and for n=2, the pH values were of about 10.35 and 11.28 for the deaerated and the aerated media, respectively. According to the values indicated in tables 2 and 3, it was derived that the Zn coatings resisted better against corrosion in the deaerated 0.1 M NaCl solutions, where the local pH was of 10.35.

However, the elimination of oxygen from the solution did not seem to affect the shape of the EIS curves and the resistance increased to 270 $\Omega.cm^{-2}$ (Table 4). The oxygen reduction seemed thus dependent on the relaxation time and influenced the overall current. The electrochemical parameters obtained from the D.C. curves in a 0.1M NaCl solution seemed to be in agreement with the results obtained from the EIS technique. Between pH 9 and 10 the corrosion products were relatively stable and behaved as a protective layer. Further alkalinisation of the solution brought about dissolution of the corrosion products, which in turn increased the corrosion current.

### 3. 2. Coating behaviour in 1M NaCl, 1M NaOH and rain water

The impedance diagrams at the corrosion potential were studied in 1 M NaCl, 1M NaOH and in natural rain water, with the EIS curves depicted in Figure 9 and the results gathered in Table 5. It was noted that the Zn coatings showed adequate resistance in both NaOH and rain water. Indeed, the $R_{ct}$ values obtained in NaOH were seven times higher than those obtained in NaCl. On the other hand, the Bode diagram showed only one time constant between 10 and 1Hz. The same behaviour was already observed on galvanised coatings [22] and was explained by the stability of the formed corrosion products. In rain water and NaOH the corrosion products are known to be primarily of $Zn_5(CO_3)_2(OH)_6$ and $Zn(OH)_2$, respectively [23,24]. In NaCl however, different intermediate species like as $ZnOH^-$, $ZnOH$, $Zn(OH)_2^-$, $Zn(OH)_3^-$ [4,8] appear prior to the formation of $ZnO$, $Zn(OH)_2$, $Zn_5(OH)_8Cl_2.H_2O$, which are more stable. Therefore, the coatings behaved better in NaOH and rain water than in NaCl.

### 4.- SUMMARY AND CONCLUSIONS

Using the EIS technique together with D.C. polarisation tests, it was shown that the dissolution anodic reactions did not reach the underlying steel. The low concentrations of NaCl (<1M) were found not to be an adequate medium to reveal (to initiate) the weakest processes and consequently to confirm the porous state of an EP steel. The nucleation and initiation of slow reactions required a very low frequency sweep (lower than 100 mHz). Conversely, the low frequencies required long response times, which may modify the surface state of the electrode. However, the Zn coatings were however shown to provide adequate corrosion protection to the steel in NaOH and natural rain water, and the effect of pores of this size was not particularly underlined.

### 5.- REFERENCES

**Table 1.- Average concentration in of natural rain water.**

| | | | Concentration (mg L$^{-1}$) | | | | | |
|---|---|---|---|---|---|---|---|---|
| pH | hardness | mineralization | Ca$^{2+}$ | Na$^+$ | K$^+$ | SO$_4^{2-}$ | Cl$^-$ | NO$_3^-$ |
| 7.1 | 65.00 | 80.00 | 20.00 | 1.50 | 1.00 | 16.75 | 10.50 | 8.00 |

**Table 2.- Electrochemical parameters obtained in different NaCl solutions and pH of the etched Zn coatings.**

| | $E_{corr}$ (mV/SCE) | $j_{corr}$ (µA cm$^{-2}$) | βa (10$^{-3}$ V/dec) | $R_p$ (Ω cm²) | $V_{corr}$ (mm y$^{-1}$) |
|---|---|---|---|---|---|
| 0.1 M–pH solution | -1020 | 99 | 74 | 283 | 1.47 |
| 0.5 M–pH solution | -970 | 146 | 87 | 223 | 2.10 |
| 1 M – pH solution | -964 | 170 | 59 | 135 | 2.53 |
| 0.1 M – pH = 5 | -1010 | 101 | 76 | 290 | 1.51 |
| 0.1 M – pH = 10 | -1015 | 129 | 96 | 246 | 0.27 |
| 0.1 M – pH = 11 | -1000 | 165 | 255 | 39 | 0.96 |

**Table 3.- EIS parameters obtained in different NaCl solutions and pH of the etched Zn coatings.**

| | $R_{ct}$ (Ω cm²) | $f_{max}$ (Hz) | CPE$_{dl}$ (µF cm$^{-2}$ S$^{(1-\alpha)}$) | α | R' (Ω cm²) | $f_{max}$ (Hz) | α' | CPE' (F cm$^{-2}$ S$^{(1-\alpha')}$) |
|---|---|---|---|---|---|---|---|---|
| 0.1M–pH solution 6.6 | 195 | *127* | 34.12 | 0.75 | -58 | 0.3 | 0.72 | -0.75 |
| 0.5M- pH solution 6.6 | 153 | 127 | 74.12 | 0.67 | -10 | 0.3 | 0.66 | -6.65 |
| 1 M – pH solution 6.7 | 108 | 115 | 85.90 | 0.71 | -18 | 9.0 | 0.70 | -0.16 |
| 0.1 M – pH = 5 | 180 | 127 | 30.50 | 0.78 | -58 | 0.3 | 0.73 | -0.52 |
| 0.1 M – pH= 10 | 179 | 204 | 18.30 | 0.79 | -66 | 0.3 | 0.75 | -0.42 |
| 0.1 M – pH = 11 | 142 | 127 | 24.60 | 0.85 | -85 | 0.5 | 0.79 | -0.14 |

**Table 4.- EIS parameters obtained in aerated and aerated 1M NaCl solution (pH =6.7) of the etched Zn coatings .**

|  | $R_{ct}$ ($\Omega$ cm$^2$) | $f_{max}$ (Hz) | $CPE_{dl}$ ($\mu$F cm$^{-2}$ S$^{(1-\alpha)}$) | $\alpha$ | R' ($\Omega$ cm$^2$) | $f'_{max}$ (Hz) | $\alpha$' | CPE' (F cm$^{-2}$ S$^{(1-\alpha')}$) |
|---|---|---|---|---|---|---|---|---|
| **Aerated** | 108 | 115 | 85.90 | 0.71 | -18 | 9.0 | 0.70 | -0.16 |
| **De-aerated** | 270 | 7 | 248.6 | 0.70 | 150 | 1.1 | 0.69 | 0.18 |

**Table 5.- EIS parameters in NaCl, NaOH and rain water at room temperature of the etched Zn coatings.**

|  | $R_{ct}$ ($\Omega$ cm$^2$) | $f_{max}$ (Hz) | $CPE_{dl}$ ($\mu$F cm$^{-2}$ S$^{(1-\alpha)}$) | $\alpha$ | R' ($\Omega$ cm$^2$) | $f'_{max}$ (Hz) | $\alpha$' | CPE' (F cm$^{-2}$ S$^{(1-\alpha')}$) |
|---|---|---|---|---|---|---|---|---|
| **1M NaCl** | 108 | 115 | 85.90 | 0.71 | -18 | 9.0 | 0.70 | -0.16 |
| **1M NaOH** | 734 | 0.1 | 390.00 | 0.66 | ---- | ----- | ----- | ----- |
| **Rain water** | 138 | 204.0 | 170.00 | 0.77 | -48 | 0.1 | 0.75 | -0.7 |

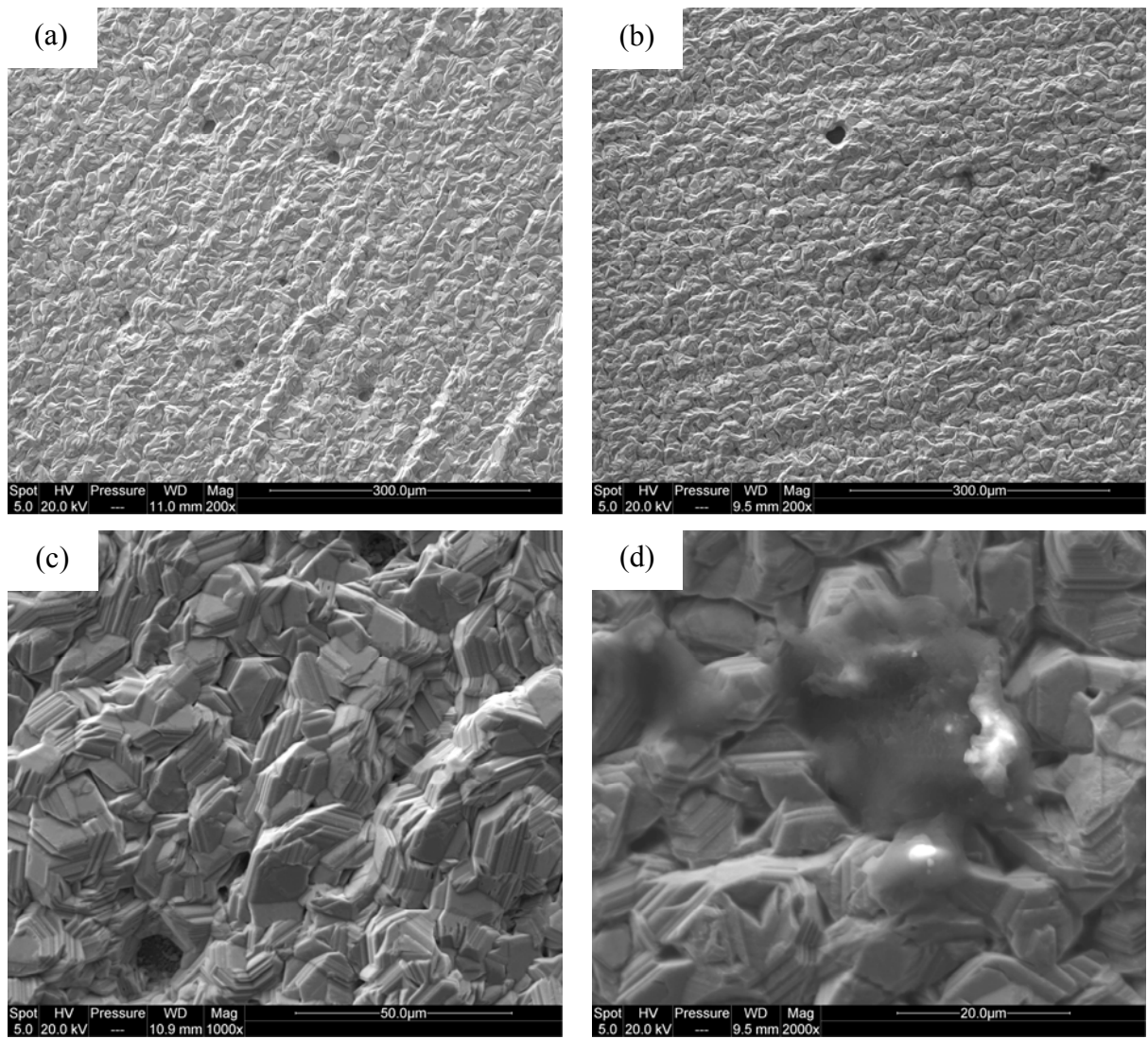

**Figure 1.- SEM images of the (a) EP Zn coating after etching in 10 vol% HCl showing the pores, (b) corrosion onset occurs at the pores, in particular at their base (c). Detail of a pore filled in with corrosion products (d).**

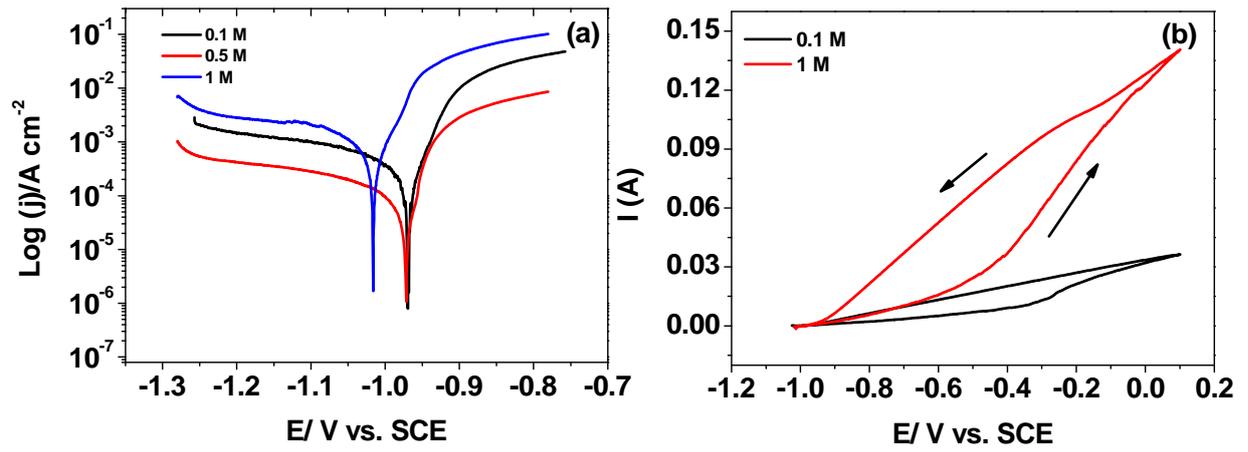

**Figure 2.- (a) Polarisation cathodic curves and (b) cyclic polarisation curves in NaCl solution.**

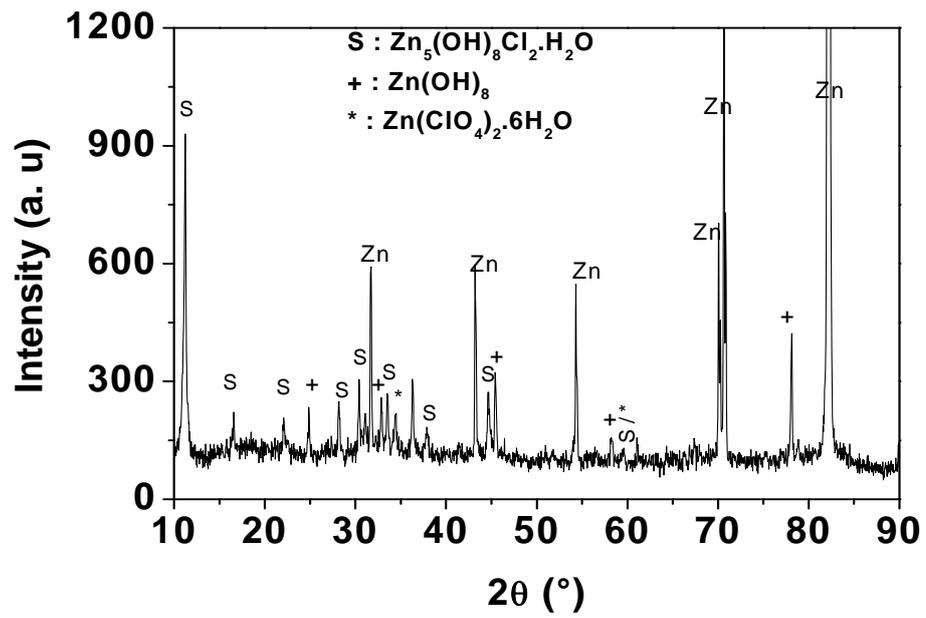

**Figure 3.- XRD typical patterns of the etched Zn coatings after polarisation in 1M NaCl.**

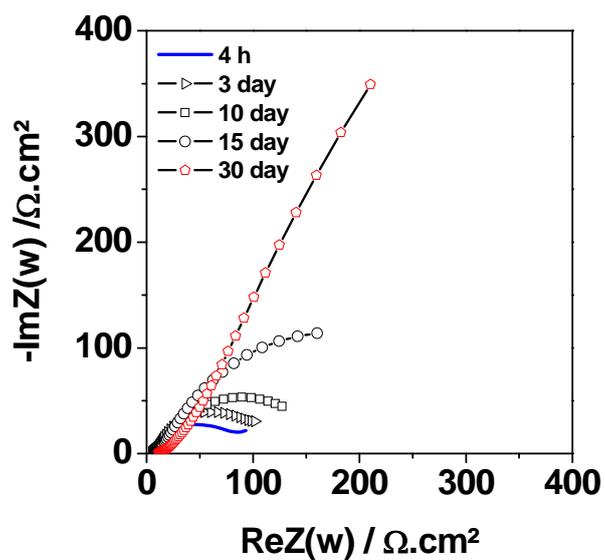

**Figure 4.- EIS diagrams of the unetched Zn coatings recorded in the 0.1M NaCl solutions as function of immersion time.**

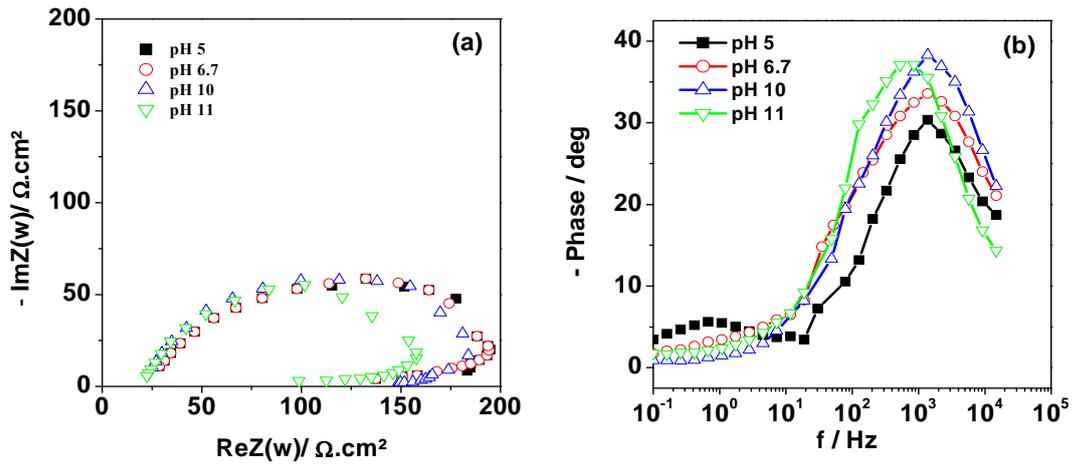

**Figure 5.- EIS Nyquist (a) and Bode (b) spectra of the unetched Zn coatings in 0.1M NaCl at different pH.**

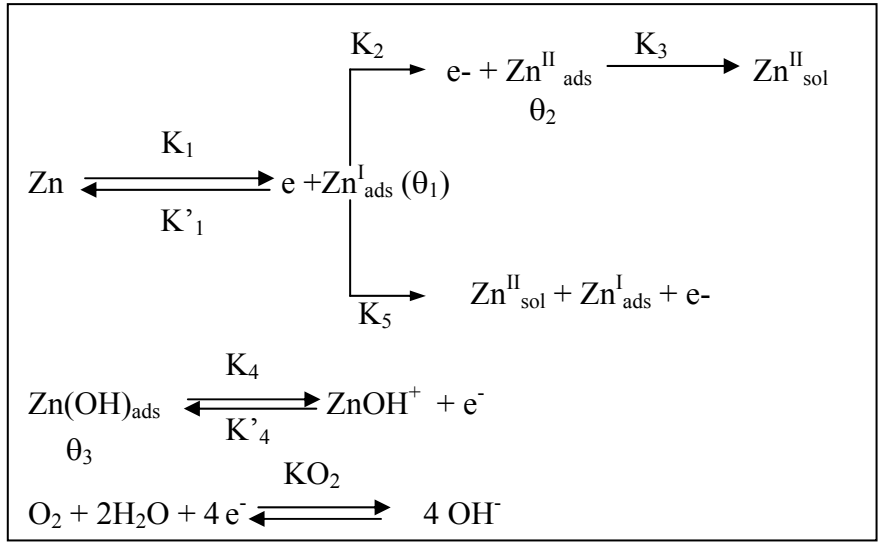

**Figure 6.- Reaction scheme for zinc dissolution in aerated NaCl solution.**

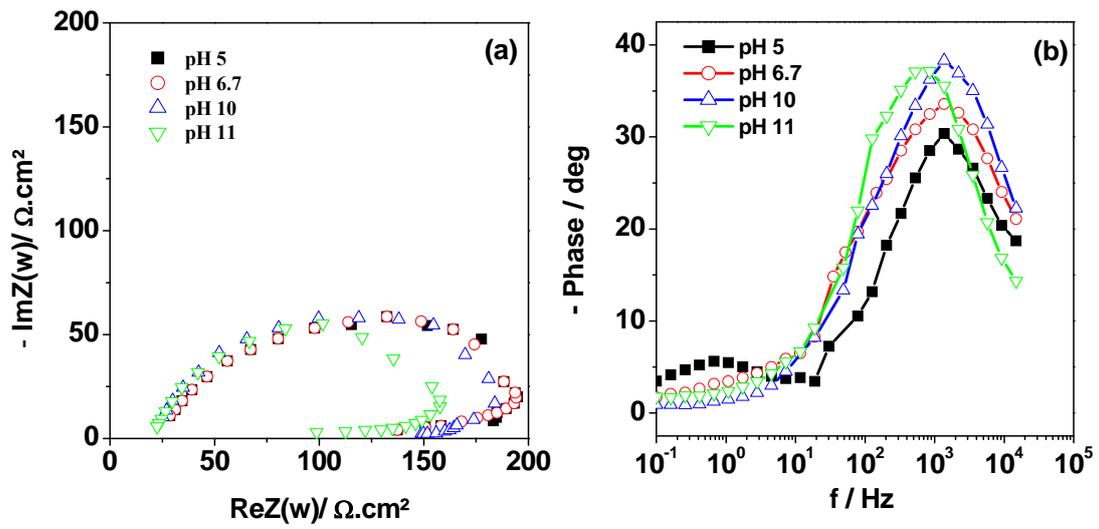

**Figure 7.-** EIS Nyquist (a) and Bode (b) spectra of the unetched Zn coatings in 0.1M NaCl at different pH.

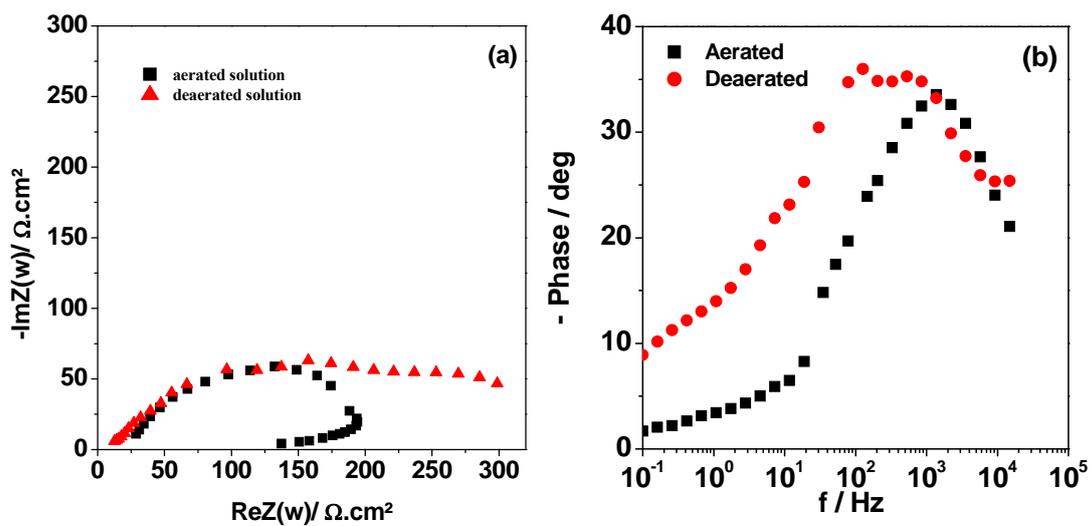

**Figure 8.- EIS Nyquist (a) and Bode (b) spectra of the unetched Zn coatings in 0.1M NaCl in aerated and de-aerated conditions.**

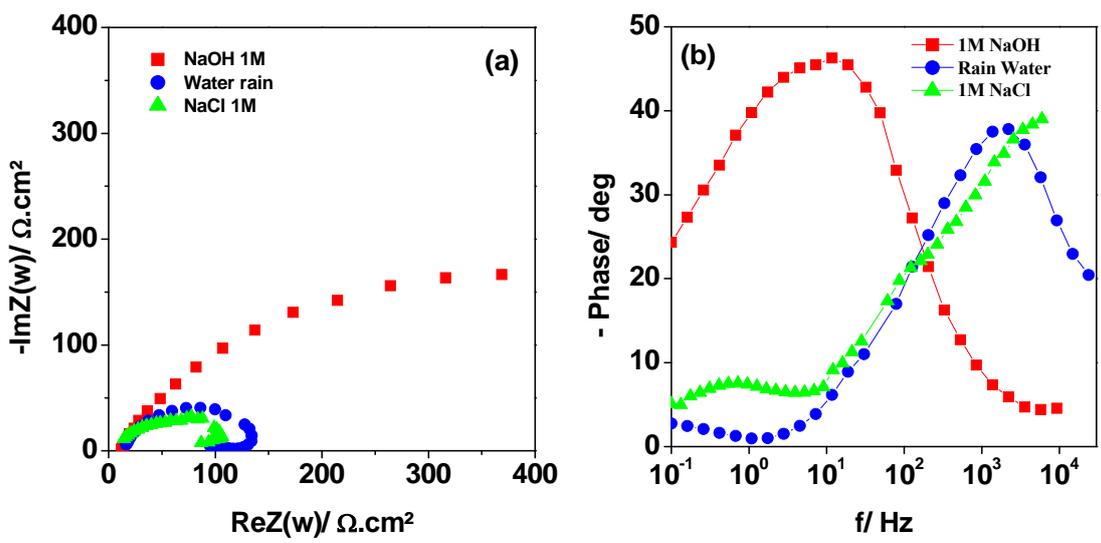

**Figure 9.-** EIS Nyquist (a) and Bode (b) spectra of the unetched Zn coatings in different media (NaCl, NaOH and rain water).